\documentclass[twocolumn,noshowpacs,preprintnumbers,amsmath,amssymb]{revtex4}

\usepackage{verbatim}
\usepackage{wasysym}

\usepackage{graphicx}
\usepackage{dcolumn}
\usepackage{bm}
\usepackage{wrapfig}

\begin{document}

\title{Growth and reductive transformation of a gold shell around pyramidal cadmium selenide nanocrystals}

\author{Michaela Meyns}
\author{Neus G. Bastus}
\author{Yuxue Cai}
\author{Andreas Kornowski}
\affiliation{Institute of Physical Chemistry, University of Hamburg, 20146 Hamburg, Germany}
\author{Beatriz H. Juarez}
\affiliation{IMDEA Nanoscience, 28049 Madrid, Spain}
\author{Horst Weller}
\author{Christian Klinke}
\email{klinke@chemie.uni-hamburg.de}
\affiliation{Institute of Physical Chemistry, University of Hamburg, 20146 Hamburg, Germany}

\begin{abstract} 

We report the growth of an unstable shell-like gold structure around dihexagonal pyramidal CdSe nanocrystals in organic solution and the structural transformation to spherical domains by two means: i) electron beam irradiation (in situ) and (ii) addition of a strong reducing agent during synthesis. By varying the conditions of gold deposition, such as ligands present or the geometry of the CdSe nanocrystals, we were able to tune the gold domain size between 1.4 nm to 3.9 nm and gain important information on the role of surface chemistry in hetero nanoparticle synthesis and seed reactivity, both of which are crucial points regarding the chemical design of new materials for photocatalysis and optoelectronic applications. 

\end{abstract}

\maketitle

Recent developments in materials physics and chemistry allow the synthesis of colloidal nanocrystals (NCs) with unique optical and electrical properties \cite{1,2}. However, it remains a grand challenge to prepare multi-component or hybrid structures in which two or more NC domains of different materials with individually tailored properties and controlled composition are integrated into one nanostructure in core-shell or oligomer-type configurations. These new hybrid NCs (HNCs) with a higher level of structural complexity are expected to exhibit modified physicochemical properties \cite{3,4} and provide systems which allow site-specific functionalization with biomolecules \cite{5,6} and their use in photocatalysis and optoelectronic applications. The most common strategy for the synthesis of these HNCs is the seed-mediated method in which one material is used as seed particle and a different material is grown around (core-shell HNCs \cite{7,8}) or forming domains on their surface (oligomer-like HNCs). Examples of the latter case are metal-semiconductor heterodimers made of spherical domains \cite{9,10,11,12}, dumbbell-like \cite{13,14}, peanut-like \cite{15} and matchstick-shaped \cite{13,16,17} structures. Several reviews on heterostructured nanocrystals summarize the progress and the perspectives on this emerging field \cite{3,4}. Additionally, these systems are unique platforms to study nucleation, growth, dissolution and reshaping of HNCs. In this context, Banin's group reported how dumbbell-like structures, formed by preferential nucleation of gold on crystal facets with high reactivity, evolved into a hetero structure with only one end covered due to a ripening process \cite{13,17} while a recent publication of Manna and coworkers investigated the structural and morphological evolution of as-synthesized CdSe-Au HNCs subjected to thermal annealing \cite{18}. The reactivities of crystal facets are closely associated to particle geometry and the degree of surface passivation. Thus, the use of NCs with a different topology as substrates for the Au growth can provide new insights not only into the formation of hybrid structures but also to the role of the NCs ligand density on the final hybrid structure.

\begin{figure}[htbp]
  \centering
  \includegraphics[width=0.45\textwidth]{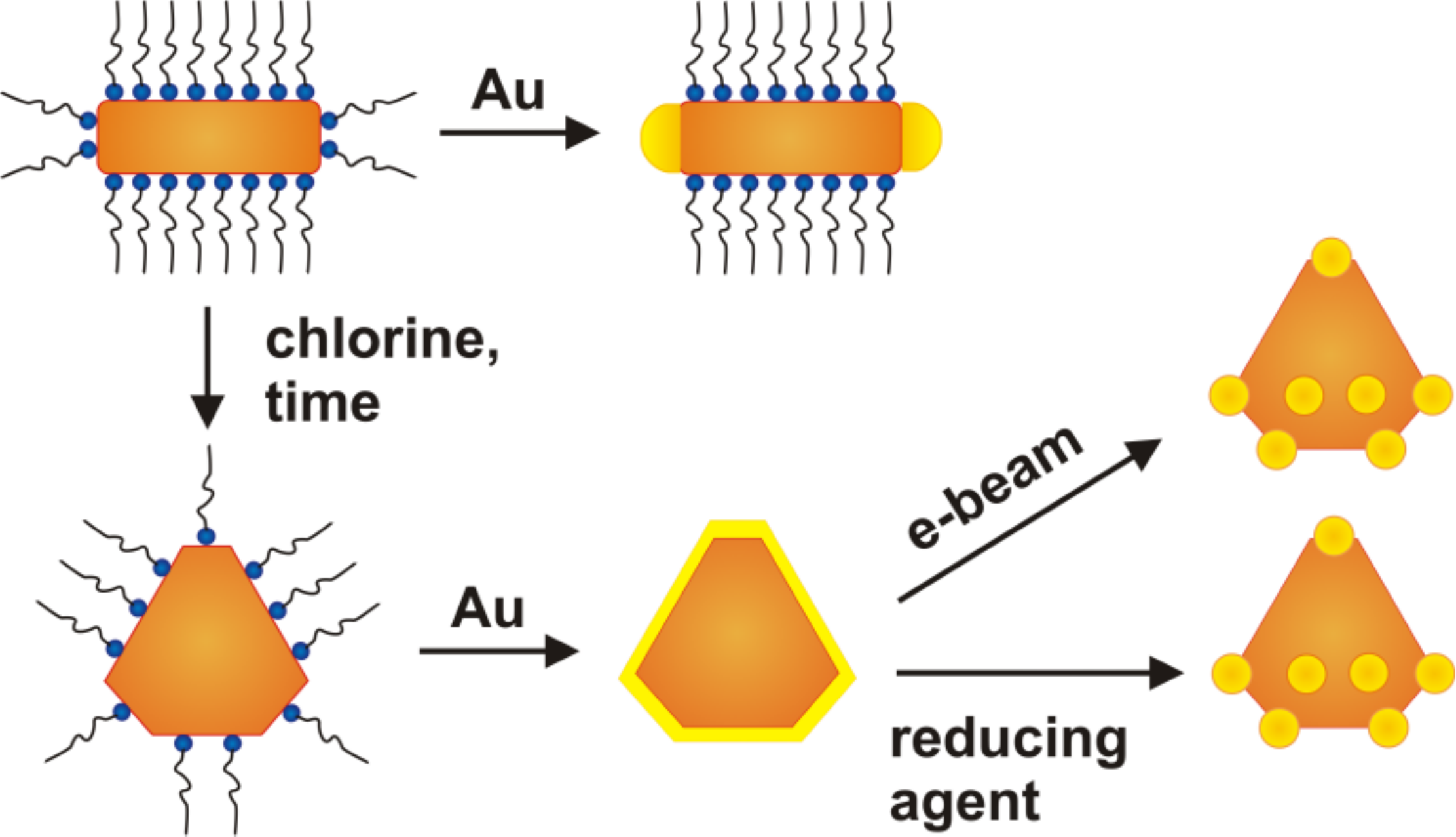}
  \caption{\textit{Representative scheme of the Au growth process onto CdSe NCs. The incubation of CdSe NCs with the Au organometallic complexes led to the formation of Au dots (top) or an instable Au shell (bottom) depending on CdSe geometry and ligand density. The sample irradiation or the addition of a strong reducing agent resulted in the nucleation of gold dots on top of the most reactive sites. The size of the final Au dots depends on the amount of gold precursor added and the ligand used.}}
\end{figure}

In this context, we report here the formation of a gold-shell structure around dihexagonal CdSe pyramids obtained upon ripening of CdSe nanorods in the presence of chlorine containing compounds \cite{19} which evolved to gold dots under electron beam irradiation (as illustrated in Fig. 1).

\begin{figure}[htbp]
  \centering
  \includegraphics[width=0.45\textwidth]{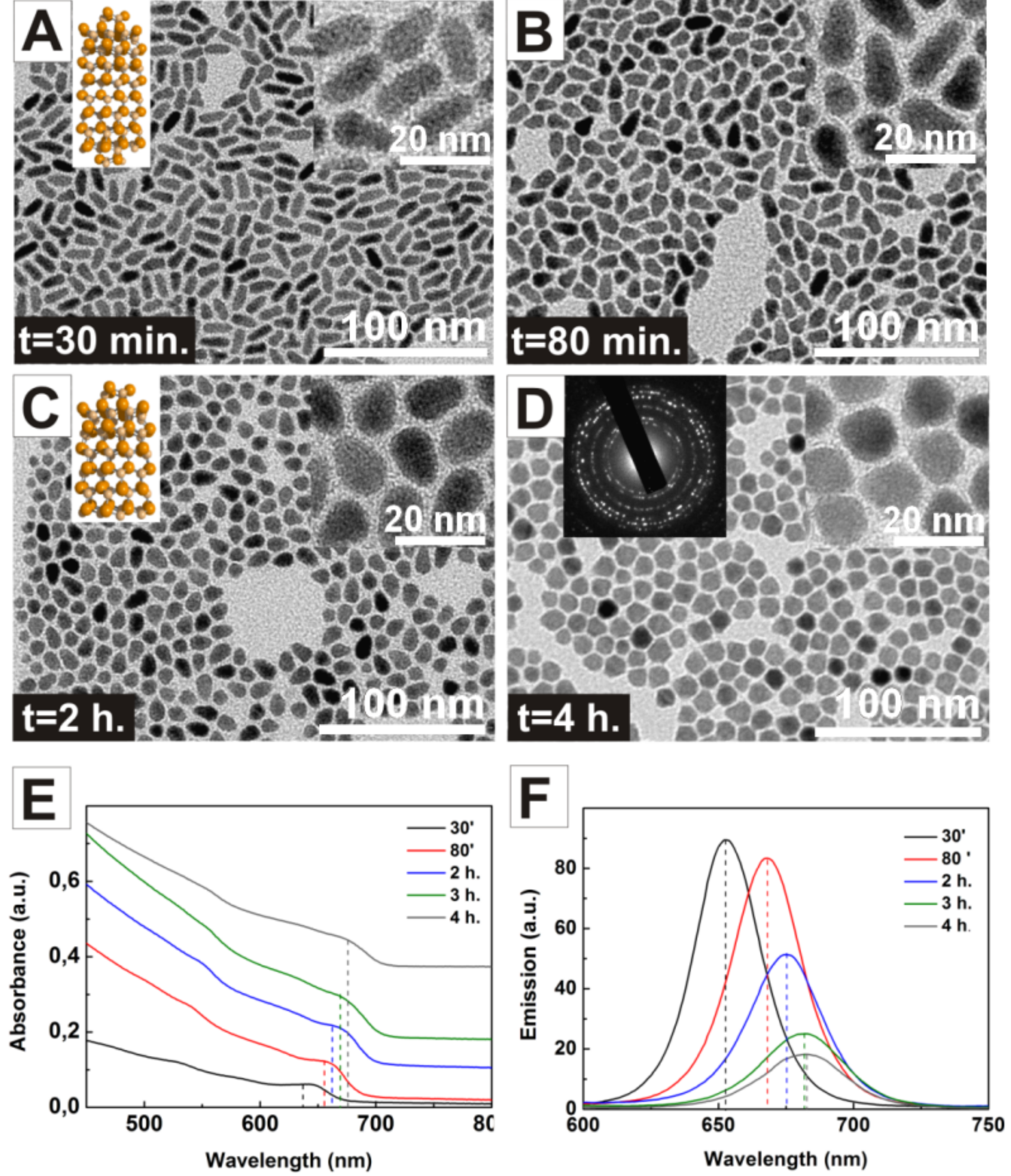}
  \caption{\textit{(A) TEM images of morphological CdSe NCs reshaping from rods to pyramids (30 min, 80 min, 2 h and 4 h). High resolution images of particles are visible in the right insets, while schematic illustrations of transformational stages at 30 min and 2 h and an electron diffraction pattern of the final nanocrystals after 4 h are visible in left insets. The optical absorption and emission spectra are displayed in (E) and (F) respectively. The absorption spectrum red-shifts and the emission decreases with time as a result of particle growth and reshaping.}}
\end{figure}

The seed particles used in our experiments were dihexagonal pyramidal CdSe NCs. With their wurtzite structure and complex geometry they intrinsically provide a high number of reactive sites. Their synthesis was carried out by introducing adaptations to a method previously published by our group \cite{19,20}. Briefly, 10 $\mu$L of 1,2-dichloroethane (DCE) were injected into a solution of CdO precursor complexed by octadecylphosphonic acid (ODPA) in trioctylphospine oxide (TOPO). The injection of Se dissolved in trioctylphosphine (TOP) promotes the formation of CdSe nuclei that grow along the c-axis of the wurtzite crystal forming rod-like NCs which then undergo a structural transformation from rods to pyramidal NCs during 4 hours of reaction. This reshaping is induced by the presence of DCE which modifies the Cd-ODPA complex and/or destabilizes the ligand environment of the rod-shaped particles and boosts an intraparticle ripening.19 Particle morphology and structural transformation from etched rods to hexagonal-pyramidal NCs was monitored by TEM (Fig. 2A). At short reaction times ($<$ 30 min), we found rod-like NPs, which grew preferentially along the c-axis with alternating close-packed layers of Cd and Se. After 2 hours of reaction these rods undergo a morphological transformation consisting of i) an etching of the more reactive (000-1) facets, ii) shaping of the facets and iii) a decrease of c-axis length. At the last stages of the reaction (4 hours) highly crystalline pyramidal NCs with 12.6 $\pm$ 1.3 nm in length (c-axis) and 11.2 $\pm$ 1.6 nm in diameter were obtained. The structural changes were also monitored by means of absorption and emission spectroscopy at different stages of evolution. In the corresponding spectra depicted in Fig. 2B-C, a red shift is visible from earlier towards later stages. The first transition band shifts from a wavelength of 648 nm to 673 nm and well-defined absorption peaks (especially the first transition) become smoother and closer to the bulk response as typical for pyramidal CdSe particles.

\begin{figure}[htbp]
  \centering
  \includegraphics[width=0.45\textwidth]{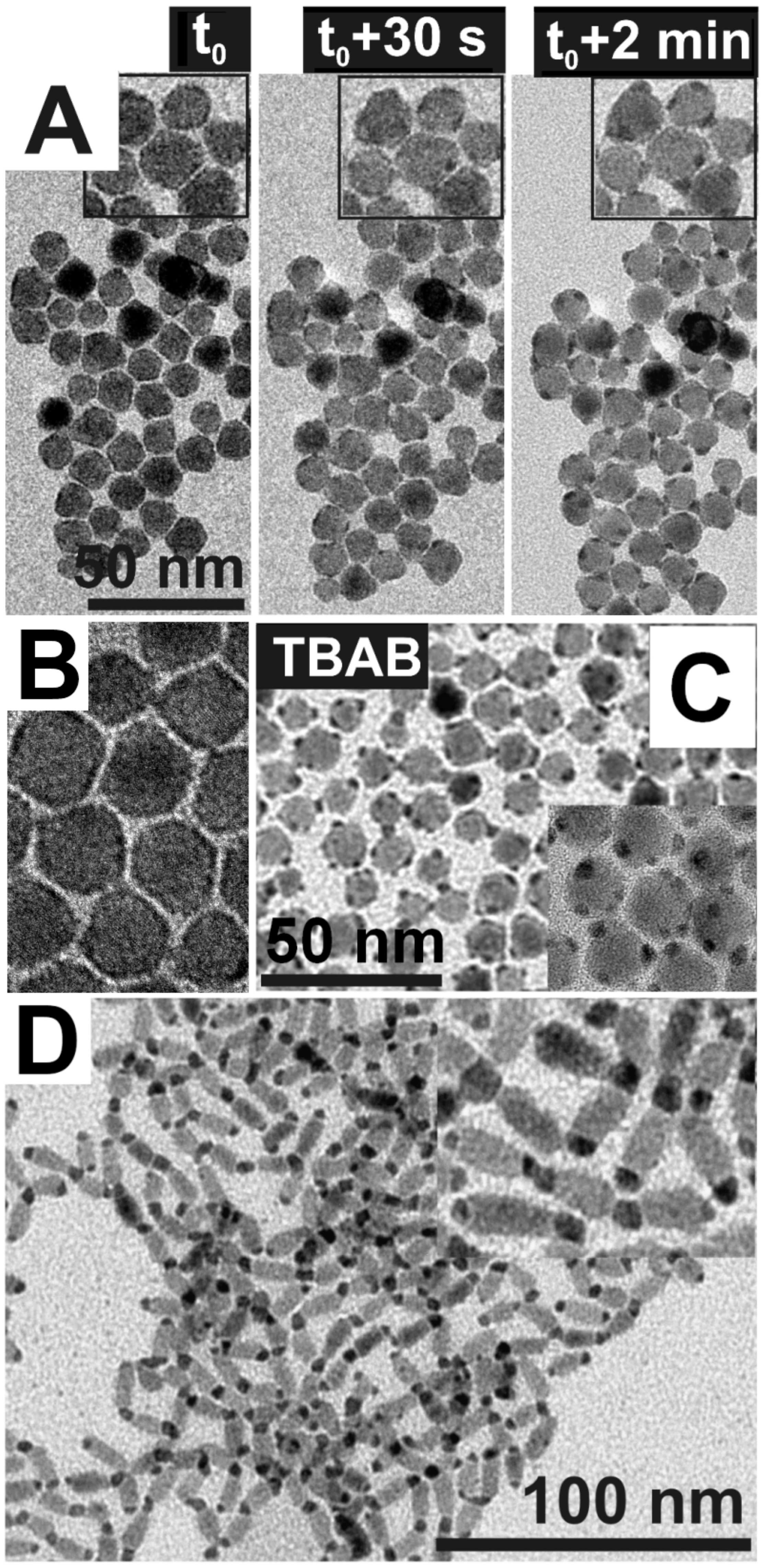}
  \caption{\textit{(A) Images taken out of a video stream obtained during TEM inspection of a sample incubated with Au-DTAB/OA but without TBAB (CdSe/Au: 2.88). (B) Enlarged image of CdSe NCs with Au-shell. (C) TEM images of CdSe NCs after their incubation with Au-DTAB/OA and further addition of TBAB-DTAB solution (Au dot diameter: 3.9 $\pm$ 1.1 nm). (D) In contrast to CdSe pyramids incubation of CdSe rods with Au precursor led directly to the heterogeneous nucleation of Au dots on the tips of the rods.}}
\end{figure}

In order to deposit gold on the surface of pyramidal NCs, we incubated them with a solution of gold(III)-chloride (AuCl$_{3}$), dodecyltrimethylammonium bromide (DTAB) and oleylamine (OA) in toluene. In this study, no extra reducing agent was added and the organic stabilizers present in the solution serve as mild reducing agents \cite{13}. TEM images show thin shell-like deposits with higher contrast around the CdSe NCs (Fig. 3A), attributed to gold. Inspecting the morphology of the hybrid particles, an interesting phenomenon was visible when comparing TEM images of one spot on the grid taken immediately after focusing and approximately 30 seconds later. During sample irradiation by the electron beam (constant beam intensity of $\sim$ 8.5 pA/cm$^{2}$), the thin shell evolved into dot-like deposits with high contrast (Fig. 3A). Au preferably migrated to vertices, leading to Au particle formation by minimizing the contact to CdSe NCs and additionally its own surface. Attempts to characterize the Au shell by XRD were made but, as no Au peak appeared, the nature of the shell appears to be non-crystalline. The dimensions of the CdSe pyramids before and after Au growth are not modified, indicating that seed particles are not oxidized in such a degree as to assume that they act as a sacrificial material for the Au growth.

\begin{figure}[htbp]
  \centering
  \includegraphics[width=0.45\textwidth]{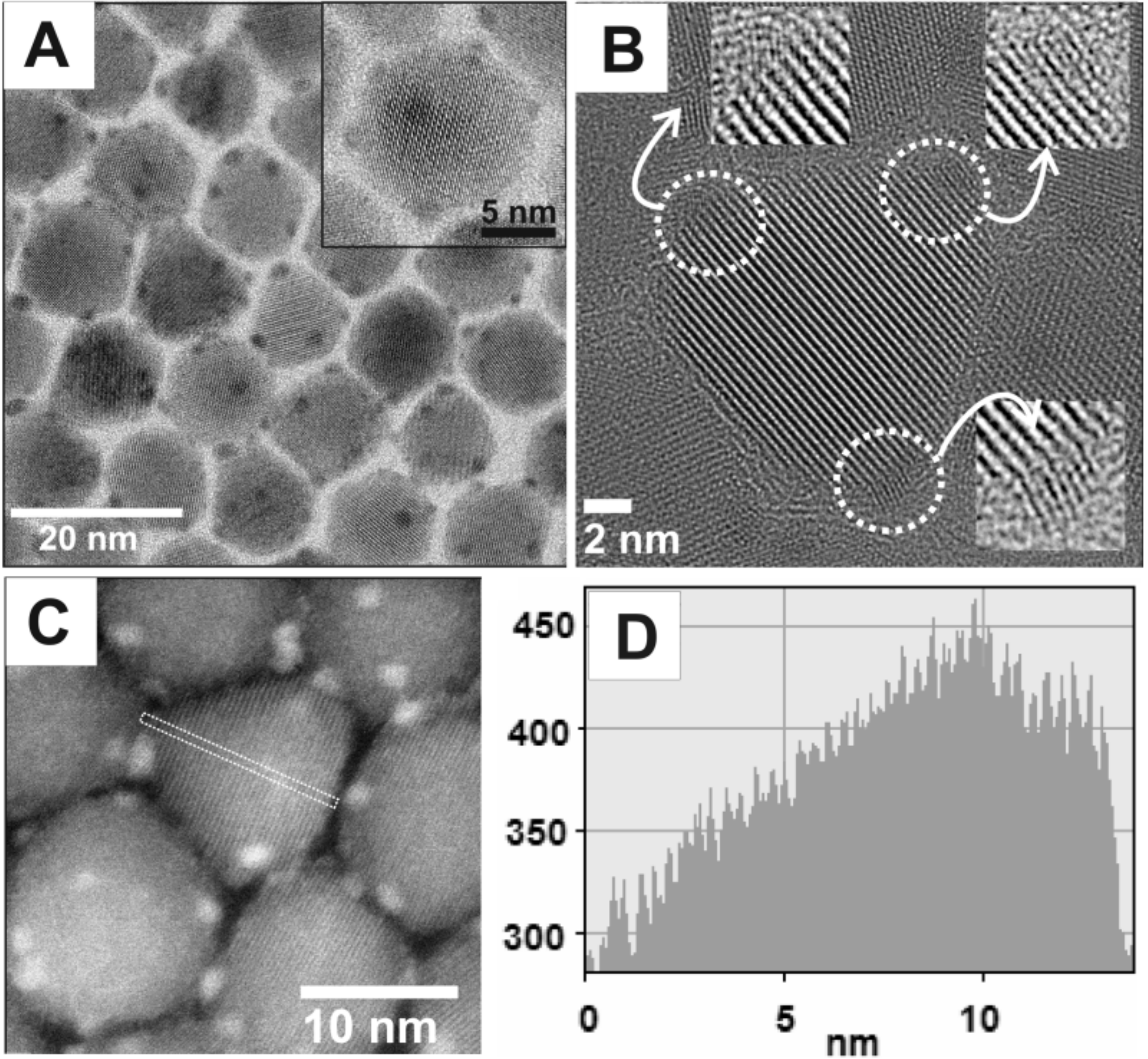}
  \caption{\textit{Incubation of CdSe with Au-DTAB/DDT (Au/CdSe: 1.92) led to Au clusters of 1.4 $\pm$ 0.3 nm positioned at apexes and nearly all vertices of the CdSe pyramids (A, inset A: top view of a pyramid). HR-TEM image B shows crystal lattices of CdSe (0002) and three gold deposits (left (311), apex (111), bottom (200)) and enlargements of the interfaces. A particle profile (D) derived from STEM measurements (C) illustrates the comparatively high curvature and thus increased reactivity of vertices along the sides of the pyramids.}}
\end{figure}

In order to determine whether this Au-shell formation was characteristic of pyramidal NCs, we did a control experiment with CdSe rods. CdSe rods were synthesized by the same method as the pyramidal particles (including the presence of DCE) except from the fact that the particles were grown for only 30 min. 

In contrast to the results obtained with the pyramids, the incubation of the CdSe rods with Au precursor led to the heterogeneous nucleation of Au dots on the tips of the rods (Fig. 3D). Once deposited, they were stable and did neither move from tips to tips nor from tips to lateral facets under electron beam irradiation. From this, two different questions arose: i) which conditions favored the formation of the shell and ii) which effect caused the migration of the Au shell under the e-beam. In previous reports, several factors responsible for the preferential deposition of a second material onto (semiconductor) nanorods have been identified, related with both, the interface between the materials (lattice mismatch, interfacial energies) and the surface of the rod component (surface defects, different ligand densites on different facets) \cite{13,14,21}. Considering that the reshaping from CdSe rods to pyramids is related with a displacement/ destabilization of the NC ligand environment, it seems plausible to consider that the ligand density on the surface of the pyramidal CdSe NCs is lower than in the case of the rods. Consequently, the whole surface of the CdSe pyramid is accessible for the Au precursor to be adsorbed/absorbed, leading to the formation of an Au shell which is thermodynamically favored over constrained growth. In the case of the rods, Au growth takes place only at their tips, in agreement with the hypothesis that Au nucleation takes place on areas with a lower concentration of ligands. Similar results were reported by Wetz and coworkers on the nucleation of Au dots on Co nanorods, where they showed the possibility to control Au dot growth (tip of whole body) through the judicious choice of metal precursor and surface ligand concentration \cite{22}.

This ligand-depleted Au nucleation hypothesis fits well with several experimental findings: i) the stability of the CdSe NCs decreases as the shape transformation is taking place, i.e. rod-shaped CdSe NCs are more stable than pyramidal-shaped ones and ii) CdSe pyramidal NCs can be directly attached to carbon nanotubes (CNT) while rod-shaped NCs cannot, i.e. the lower passivation of the CdSe pyramids allow CNTs to act as a "ligand" for the NCs \cite{19,20}.

The unexpected instability of the Au-shell on pyramidal CdSe NCs under the electron beam can be understood either in terms of a thermal annealing effect or due to an additional reduction of Au species on the CdSe NC surfaces by electrons from the beam \cite{23}. In order to elucidate which of these phenomena caused the migration of Au atoms on the CdSe surface we performed additional experiments by increasing the temperature of the CdSe solution up to 80$^{\circ}$C before the injection of the Au-OA precursor (above precursor decomposition temperature \cite{24}). After several days of reaction during which several aliquots were taken, the samples were characterized obtaining similar results in all cases, namely, the formation of an unstable Au-shell structure similar to the one obtained at R.T. These results suggest that higher incubation temperatures do not prevent the evolution of the Au shell and hence, that the shell instability could rather be due to an incomplete Au reduction. In order to prove this assumption, we used tetra-n-butylammonium borohydride (TBAB) as a strong reducing agent. This compound, used in the organic synthesis of Au NPs \cite{25}, is less reactive and less polar than NaBH$_{4}$ but considerably stronger than amines. Here, a freshly prepared TBAB-DTAB (1:1, toluene) solution was added to a dispersion of CdSe NCs previously incubated with Au precursor. Fig. 3B shows the morphological characterization of hybrid structures obtained after 20 min of reaction. In comparison with those samples obtained without TBAB (Fig. 3A), the use of the strong reducing agent led to the formation of well-defined and stable Au dots with an average size of 3.9 $\pm$ 1.1 nm which grew onto the CdSe NCs, in detail, on the sites with high curvature, especially the apex. 

Since the final properties of the HNC depend on the dimensions of the material grown, we studied the possibility to tune the size of the Au dots by changing the concentration of Au (Au to CdSe ratio: molar amount Au/(optical density*volume CdSe dispersion)). By using half of the initial amount of Au, for example, we obtained Au dots of 2.5 nm in diameter in which the number of dots was comparatively lower (around two or three per particle in contrast to at least 4). Additionally, inspired by a method for the organic synthesis of Au NPs published by Jana and coworkers \cite{25} we tried to control the size of the Au dots by changing the ligand used on the precursor solution from OA to dodecanethiol (DDT). High resolution TEM and STEM images (Fig. 4 A-C) reveal how the exchange of OA for DDT (while maintaining the original Au/CdSe ratio) led to the formation of ultra-small Au clusters with a diameter of 1.4 $\pm$ 0.3 nm placed on vertices of the dihexagonal CdSe pyramids. A profile derived from a Scanning Transmission Electron Microscope (STEM) measurement illustrates the geometry of a dihexagonal CdSe pyramid (Fig. 4D). The structure provides relatively sharp angles which are highly reactive sites for the nucleation of gold structures and the further formation of rather spherical NCs. 

Selected area energy-dispersive X-ray spectroscopy (EDX) analysis performed on the obtained HNCs using OA as a ligand, revealed an increase of the relative Au weight amount (from 15.3\% to 30.1\%) as the Au/CdSe ratio increases (from 0.96 to 1.92). In parallel, analysis performed on HNCs with DDT as a ligand (Au/CdSe: 1.92) indicated a decrease of the Au content (8.2\%) as well as the presence of sulfur. UV-Vis spectra recorded from CdSe HNCs show similar tendencies as reported for other directly attached metal-semiconductor hybrid systems (smoothed absorption features and quenched emission) \cite{13,26}. Important information about the crystalline interface between CdSe and Au can be obtained at even higher magnification (Fig. 4B). The lattice fringes of the Au partially merge into those of CdSe. To equilibrate stress due to the lattice mismatch between Au and CdSe (5\% a-axes, 58\%, c-axes) edge dislocations are present in the Au clusters. The direct contact between both materials proofs that structural changes on the surface during the transformation from rods to pyramids do not interfere with Au nucleation onto surface layers of CdSe. The fact that lattice distances in gold deposits on the apex of the pyramids (with their (0002) facet visible) indicate a (111) facet matches with earlier reports on lattice orientation of Au on CdSe rods \cite{13}. Further investigations need to be carried out in order to obtain reliable information about preferred orientations between Au and CdSe lattices in other positions, as relative orientations of the components in the path of the electron beam are difficult to be obtained from single 2D TEM images. 

In conclusion, the comparison of gold deposition behavior on CdSe rods and pyramids together with the results obtained "ex-situ" by adding a strong reducing agent suggests that i) the formation of the Au shell could be correlated with the lower degree of passivation of the HNC surface and ii) the origin of the shell transformation could be due to an incomplete reduction of the gold shell. We believe that our findings may be of general utility to create other types of complex colloidal heterostructures and help to interpret structural instabilities or inhomogeneities in samples occurring after the actual synthesis. The method presented is a straightforward technique for the synthesis of nearly monodisperse CdSe-Au HNCs with quasi-epitaxial interfaces and controlled size and shape of Au deposits on the most reactive CdSe NC sites. Especially hybrid structures with cluster-sized Au domains might open new possibilities for applications in catalysis or optoelectronics.

\end{document}